\documentclass[lettersize,journal]{IEEEtran}
\usepackage{amsmath,amsfonts}
\usepackage{algorithmic}
\usepackage{algorithm}
\usepackage{array}
\usepackage[caption=false,font=normalsize,labelfont=sf,textfont=sf]{subfig}
\usepackage{textcomp}
\usepackage{stfloats}
\usepackage{url}
\usepackage{verbatim}
\usepackage{graphicx}
\usepackage{cite}
\usepackage{fontenc} 
\usepackage[utf8]{inputenc}
\begin{document}


\title{Measurement of Multilayer Coating Thickness on Interwoven Carbon-Fiber-Reinforced Polymers Using the Terahertz PHASR Scanner}

\author{{Arash Karimi, Zachery B. Harris, Erica Heller, Paul Vahey, M. Hassan Arbab, Member, \textit{IEEE}}

\thanks{This research received no external funding. The authors thank The Boeing Company for providing the samples used in this study. (\textit{corresponding author: M. Hassan Arbab}, email: hassan.arbab@stonybrook.edu)} 
\thanks{Arash Karimi, Zachery B. Harris, Erica Heller, and M. Hassan Arbab are with the Department of Biomedical Engineering, Stony Brook University, Stony Brook, NY 11794, USA }
\thanks{Paul Vahey is with Boeing Research and Technology, Seattle, WA 98108, USA}}


\IEEEpubid{0000--0000/00\$00.00~\copyright~2021 IEEE}
\maketitle
\begin{abstract}
Coating thickness measurement and inspection of defects beneath optically opaque surfaces are among the most promising commercial applications of the terahertz (THz) imaging technology. However, there are two main sources of complexity in THz spectral measurements that have resulted in reduced accuracy and high uncertainty in the determination of coating thicknesses. These factors include: 1. The need for \textit{a priori} knowledge of the complex index of refraction of each coating layer, and 2. The complex and polarization-sensitive reflectivity of samples with carbon-fiber-reinforced polymer (CFRP) substrates. In this paper, we propose a combined hardware and software solution to address these two limitations. We employ the polarimetric version of our Portable HAndheld Spectral Reflection (PHASR) Scanner and use a novel training model on time-of-flight measurements obtained by sparse deconvolution of the THz time-domain pulses. Our method does not depend on separate measurement of the index of refraction of the coating layers; rather, it relies on a physics-based linear model, trained using a small subset of the sample data. We show that thickness measurements with mean square error of about 10 $\mu$m and accuracy of more than 92\% can be achieved when the useful bandwidth of the scanner is limited to about 1.5 THz using photoconductive antenna emitters and detectors. We show that the anisotropic response of the carbon fiber bundles and the weave pattern structure of the substrate can significantly influence the accuracy of the coating measurement, and therefore a polarimetric imaging approach should be used for inspection of similar samples.     
\end{abstract}

\begin{IEEEkeywords}
Coating thickness measurement, interwoven carbon-fiber reinforced polymer (CFRP), sparse deconvolution, terahertz time-domain polarimetric imaging
\end{IEEEkeywords}

\section{Introduction}
\IEEEPARstart{T}
\noindent\uppercase{erahertz} spectroscopy is an emerging imaging technology that has promised great applications during the past few decades. Due to its non-ionizing, non-destructive, and non-contact nature \cite{hintzsche2013nonionizing, tao2020THz, sitnikov2021effects, jepsen2011terahertzreview}, THz imaging has drawn attention in biomedical imaging \cite{khani2023burn, chen2022cornea, cheon2017cancer, virk2021development,bowman2016terahertz}, security \cite{tzydynzhapov2020security, li2018study, cheng2022improved,shchepetilnikov2020new}, and non-destructive testing (NDT) for in-line quality control inspection \cite{amenabar2013introductory, yakovlev2015non, palka2016non}, among others. Since many industrial dielectric materials are transparent in the THz regime, these wavelengths have been suitable as an NDT technology for defect identification \cite{wang2022real,mei2021terahertz, dong2015nondestructive} and contamination detection \cite{cheng2022contamination, lu2022contamination,stelmaszczyk2022ultrafast}, material characterization\cite{khani2021diffuse, sun2020exploiting ,palka2016characterization, khani2021chemical, khani2022multiresolution, zhou2021effective}, and coating thickness measurement \cite{zhong2019progress, zhai2020nondestructive, lin2017measurement}. THz techniques have found applications in electronics \cite{burford2014terahertz, true2021terahertz, shur2019sub}, pharmaceutical \cite{zhong2011non, kawase2013non, moradikouchi2022terahertz}, automotive \cite{krimi2016highly, ketelsen2022thz, dong2015non}, aerospace \cite{ospald2014aeronautics, cristofani2014nondestructive, wang2019nondestructive, ryu2016nondestructive}, and many other industries.

The measurement of the thickness of multilayer coatings is an unmet need for cost and quality optimization in many advanced manufacturing industries. Traditional technologies used in aerospace and automotive sectors are eddy current, microcomputed X-ray tomography (XCT), and ultrasonic imaging \cite{zhong2019progress, yasui2005terahertz}. Eddy current-based methods are limited to metal substrates and can only measure the total coating thickness \cite{yasui2005terahertz, garcia2011non}. Ultrasonic imaging can be used to evaluate metallic and dielectric substrates, but requires direct contact with the sample and has limited resolution \cite{zhong2019progress}. 
In addition to safety considerations for X-rays, materials with low atomic numbers (e.g., carbon) show a poor contrast in this frequency regime \cite{thompson2016x}, making XCT unsuitable for many NDT applications. In other words, carbon composite materials have an attenuation coefficient in X-ray frequencies close to that of epoxy resin, making the use of XCT challenging for related applications \cite{naresh2020use}. Therefore, there is a deficiency in the performance of currently available technologies for direct coating thickness measurement on carbon fiber substrates, which can be addressed using THz time-domain spectroscopy (THz-TDS) imaging. 

\IEEEpubidadjcol

Further, polarization information of the THz light provides a deeper understanding of the structure of anisotropic samples, such as interwoven CFRP. Carbon fibers can act as imperfect wire-grid polarizers in the THz frequencies, and depending on the relative polarization of the incident THz beam, they can reflect the light almost fully or partially. 
Moreover, the uneven surface of the interwoven carbon fibers requires careful consideration of spatial information for either inspection of delamination or measurement of coating thicknesses.
Studies have shown promising results on the use of THz-TDS in CFRP structures for defect detection \cite{ospald2014aeronautics, dong2018visualization} and measurement of coating thickness \cite{su2014terahertz} in such samples. However, because of the anisotropic structure of interwoven CFRPs, there is a need for polarimetric THz imaging. 
Therefore, this need has led us to develop and calibrate the polarimetric version of our Portable HAndheld Spectral Reflection (PHASR) Scanner, using two orthogonally oriented photoconductive antenna (PCA) detectors separated by a polarizing beam splitter \cite{harris2024handheld, harris2022terahertz}. The fast THz full-spectral imaging capability of the PHASR Scanner has already been demonstrated during in vivo burn severity diagnosis \cite{khani2022accurate, khani2022supervised, osman2022deep}, corneal imaging studies \cite{virk2023design}, etc. 
Using the polarization sensitive version of the PHASR Scanner, we recently showed that polarimetric imaging of interwoven CFRP adds deeper insight into such anisotropic structures \cite{xu2024thz}.
In this work, we will demonstrate the ability of the PHASR Scanner for rapid in-line imaging of complex aerospace samples fabricated with interwoven carbon fibers in the substrates using the new polarimetric capability.

\par
Aside from a hardware solution for the THz imaging modality in NDT applications, we propose a novel algorithm to measure the thickness of multilayer coatings using a linear regression training model. Various signal processing approaches have been introduced to extract thickness of multilayer coatings. These approaches include time-of-flight analysis of THz echoes, which requires accurate knowledge of the optical properties of the layers \cite{cao2019noncontact}, and numerical fitting algorithms, which are able to extract both the thickness and the optical properties of coating layers by modeling the beam propagation in the multilayered structure \cite{krimi2016highly}. However, these methods are computationally expensive or require measurements from a calibration set to determine the complex refractive index of each coating layer. Furthermore, uneven and patterned substrates made of interwoven CFRP add additional complexity to these models. Finally, according to Fresnel's reflection law, separate echoes from each coating layer can be observed because of a difference in the refractive indices of these layers. However, having layers with similar refractive indices in THz frequencies, or thicknesses smaller than the time-of-flight associated with the full-width-at-half-maximum of the THz pulses, can increase the uncertainty of parameter extraction. If the reflected echoes are not well separated in the time domain, it would be challenging to extract both the refractive index and, consequently, the thickness of layers from the overlapped echoes. To overcome this limitation, optimization algorithms have been introduced \cite{krimi2016highly,su2014terahertz,scheller2009analyzing}. 


In this paper, first, we will present THz polarimetric imaging of interwoven CFRP substrates using the PHASR Scanner. Next, we will propose a simple model of multilayer coating layers with close refractive indices based on the information extracted from the sparse deconvolution of THZ-TDS pulses. We will show that by training a linear regression model both the effective index of refraction of the entire multilayer coating system, as well as the initial thickness of the base layer, can be extracted from the slope and y-intercept of the line. Finally, we will show that using this model we can extract the thickness of the coating layers, which were not used for the training of the model, with an accuracy better than 92\% and a mean square error of about 10 $\mu$m.


\section{Materials and Methods}
\subsection{THz Time-of-Flight Topography}
\label{THz Time-of-Flight Topography}
In multilayered structures, the interface of thin layers with different refractive indices results in Fabry-P\'erot internal reflections within the sample. The temporal separation of the reflected echoes, $\Delta t$, is related to the optical path length or the thickness of the layer, which for a normal incident beam is given by,
\begin{equation}
\label{OPL1eq}
\Delta t = \frac{2\,OPL}{c},
\end{equation}
where $c$ is the speed of light in vacuum, $OPL = n\,d,$ is the optical path length of a layer equal to half of a roundtrip, $n$ is the refractive index and $d$ is the thickness of the medium. (\ref{OPL1eq}) converts the time-of-flight of the THz echoes to the thickness of the sample, given an \textit{a priori} knowledge of the index of refraction, $n$, which may not be readily available or can be constant in heterogeneous or anisotropic samples. The coupling of the refractive index and thickness is a fundamental limitation of time-of-flight topography and can give rise to uncertainties in the determination of the coating thicknesses. 



\subsection{Sparse Deconvolution}
\label{sparse deconvolution}
The superposition of the THz time-domain reflected echos from a sample can be described by the incident beam convolved by the sample's impulse response function, which is a desired parameter to calculate in a measurement. However, extracting this impulse response function from the sample and reference measurements in the presence of noise is an inverse problem. 
This impulse response of a multilayered structure can be modeled by the finite impulse response (FIR) system, which is given in terms of a convolution sum by \cite{bourguignon2011sparse},
\begin{equation}
\label{LTIeq}
y[m]=h_0x[m]+h_1x[m-1]+h_2x[m-2]+... ,
\end{equation}
where $x[m]$ is the incident THz pulse (obtained from the reference measurement), and $y[m]$ is the output (i.e., the reflected pulse from the sample). In the presence of noise, this summation can be considered as,
\begin{equation}
\label{convolution}
y[m]=h[m]*x[m]+\epsilon[m] =\sum_{i} h[i]x[m-i] +\epsilon[m],
\end{equation}
where $h[m]$ is the impulse response of the sample, and $\epsilon[m]$ is the noise of measurement. 
Calculating the impulse response $h[m]$ in frequency-domain is an ill-posed problem, hence,
it is preferred to apply the deconvolution operation in the time domain. For this purpose, the convolution operator in (\ref{convolution}) can be converted to a matrix multiplication format,
\begin{equation}
y=Xh+\epsilon,
\label{deconvmatrixform}
\end{equation}
where rows of $X$ are time-shifted versions of the reversed (flipped) $x[m]$. Therefore, $h$ can be calculated by optimizing the following equation,
\begin{equation}
\label{gradientdescent}
\frac{1}{2}\|Xh-y\|_2^2+\lambda \|h\|_p,
\end{equation}
where the second term in (\ref{gradientdescent}) is the regularization term accounting for $\epsilon [m]$, and $\lambda$ is the regularization parameter, which is a trade-off between the fidelity to the measurement and the noise \cite{beck2009ITS}. The impulse response $h[m]$ is expected to have only few non-zero elements, or in other words to be sparse. Therefore, we can apply the sparsity constraint to (\ref{gradientdescent}) by setting the regularization term as the $\ell_0$-norm of $h$, which is the number of the non-zero elements in $h$. Nevertheless, solving (\ref{gradientdescent}) with $\ell_0$ penalty is a non-polynomial hard problem and computationally infeasible to solve \cite{natarajan1995sparse}. It was shown that replacing $\ell_0$ by $\ell_1$, which constraints the sum of absolute values of $h$, made (\ref{gradientdescent}) convex and solvable \cite{taylor1979deconvolution, zuo2013sparsedeconv, candes2008deconvolution}. It is worth noting that choosing the $\ell_p$ penalty with $p>1$ results in non-sparse deconvolution, which introduces more noise into the result \cite{yarlagadda1985fast}.
The solution to (\ref{gradientdescent}) can be obtained using the iterative shrinkage algorithm (ISA) \cite{beck2009ITS}, where the impulse response, $h$, is iteratively updated by soft thresholding the gradient descent of (\ref{gradientdescent}). ISA has been introduced for NDT applications in THz-TDS, and a detailed description of this algorithm can be found in \cite{dong2017terahertzdeconv}. In addition, sparse deconvolution has been used in ultrasound \cite{wei2009sparseotherapps}, fluorescence microscopy \cite{hugelier2016sparse}, and image processing \cite{bronstein2005blind}.


\subsection{The Polarimetric THz PHASR Scanner}
\label{The Polarimetric THz PHASR Scanner}
\begin{figure}[t]
\centering
\includegraphics[width=\linewidth]{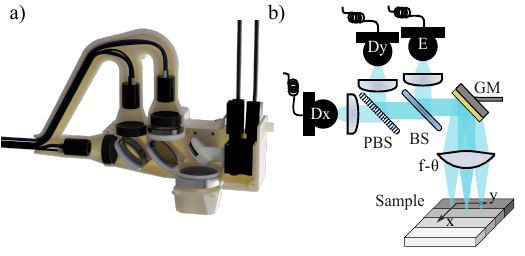}
\caption{a) Photograph of the polarimteric THz PHASR Scanner. b) Schematic of the components of the scanner with the multilayer coated CFRP sample.}
\label{scanner}
\end{figure}

Fig.\ref{scanner}(a-b) shows the photograph and schematic of the components in the polarimetric PHASR Scanner, respectively. The scanner, its polarimetric calibration, and performance characterization have been described elsewhere \cite{harris2024handheld}. It is based on a fiber-coupled photoconductive antenna (PCA) emitter and two detectors, pumped by two femtosecond fiber lasers, which have been configured to operate in the Electronically Controlled Optical Pulse Sampling (ECOPS) mode (TeraFlash smart, Toptica Photonics AG, Germany) \cite{dietz2014all}. Briefly, the THz beam is generated by the PCA emitter, labeled E in Fig.\ref{scanner}(b), at 45$^\circ$ linear polarization, collimated by a TPX lens, and reflected from a beam splitter (BS). After reflection from a motorized gimbal mirror (GM), the beam is focused perpendicularly to the sample using a custom f-$\theta$ scanning lens. Upon reflection from the sample, the beam follows the same optical path through the f-$\theta$ lens and the GM, and after passing through the BS, it is separated into two orthogonal polarizations using a polarizing beam splitter (PBS). The two orthogonal detectors, labeled Dx and Dy in Fig.\ref{scanner}(b), measure the reflected beam from the sample. In this study, an 8 mm $\times$ 8 mm area of interest (ROI) is raster scanned with a 0.25 mm $\times$ 0.25 mm pixel size. Each pixel contains a full time-domain THz pulse trace over 100 $ps$, which is averaged over 10 measurements from the same location. Similarly, a THz-TDS image from a gold-coated mirror was captured as a reference measurement. We carefully ensured that the reference and samples were positioned at the same focus distance from the f-$\theta$ lens. However, exact alignment between the time of arrival of the reference and sample pulses is not needed for the following signal processing steps. 

\section{THz Polarimetric Imaging of Interwoven CFRP Samples With Multilayered Coating}
\label{THz Imaging of CFRP With Multilayered Coating}
\begin{figure}[b]
\centering
\includegraphics[width=\linewidth]{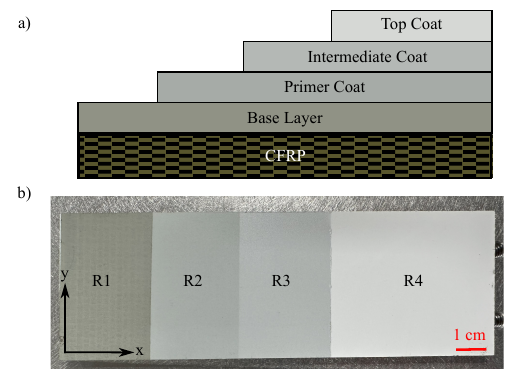}
\caption{The side view of the sample shows the cross section of an interwoven CFRP sample, coated with step-profile paint layers (a). Top view photograph of a sample with coating regions indicated (b). The x and y axis shows the orientation of the sample with respect to the lab coordinate system.}
\label{samplefig}
\end{figure}

The target samples are CFRP panels ($n$ = 23), painted with 4 layers of coating, in a step-profile pattern, shown in Fig.\ref{samplefig}(a). Fig.\ref{samplefig}(b) shows the resulting 4 regions (i.e., R1-R4) that are created on the samples by the coating layers, which include the Base Layer (BL), Primer Coat (PC), Intermediate Coat (IC), and Top Coat (TC), as shown in Fig.\ref{samplefig}. The thickness of the PC, IC, and TC layers varies from sample to sample and was provided by the manufacturer (The Boeing Company, Seattle, WA), while the thickness of BL was unknown. 
Importantly, while the individual thickness of each layer was provided through the borehole measurement technique using a Paint Borer, the first author was blinded to these values.


\begin{figure}[!t]
\centering
\includegraphics[width=\linewidth]{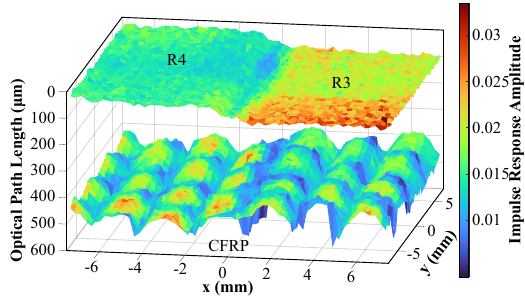}
\caption{THz topographic image of a 15 mm $\times$ 15 mm ROI over the boundary of the R3 and R4 of a sample, using the impulse response function of the pixels. The two surfaces from top to bottom indicate the relative OPL of the impulses from the air-coating interface and the CFRP substrate, respectively.} 
\label{3Dimage}
\end{figure}

A 15 mm $\times$ 15 mm scan from the boundary of the R3 and R4 area of a representative sample is shown in Fig.\ref{3Dimage}. The impulse response function of each pixel is calculated using the ISA deconvolution, described in \ref{sparse deconvolution}, which was then used to form a 3D topographic image of the sample. The impulse response of the sample has two main positive peaks, the first of which corresponds to the THz pulse reflected at the air-IC and air-TC interfaces, as indicated by the R3 and R4 surfaces in Fig.\ref{3Dimage}, respectively. The second positive impulse is from the interface between the coating layers and the interwoven CFRP substrate. Fig.\ref{3Dimage} shows the step profile in the air-coating impulse and the nonuniform structure of the interwoven CFRP substrate, which will be discussed in more detail later in this section. Crucially, the unique shape of the substrate should be well understood and taken into consideration if an accurate account of the variation in the coating layer thicknesses is desired from the analysis of interwoven CFRP samples.   

\begin{figure}[!b]
\centering
\includegraphics[width=\linewidth]{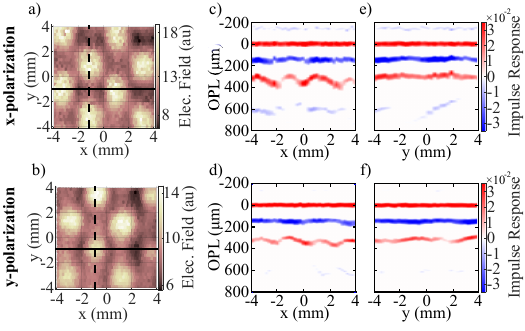}
\caption{Polarization-sensitive THz topographic scans of a coated CFRP sample. C-scan of electric field at the TOA of the coating-CFRP interface echo, measured by the (a) x-polarized and (b) y-polarized detectors. Corresponding B-scans along the horizontal lines (solid lines) constructed by the impulse response of the sample, measured by the (c) x-polarized and (d) y-polarized detectors. B-scans along the vertical lines (dashed lines) using the impulse response of the sample recorded by (e) x-polarized and (f) y-polarized detectors.}
\label{Bscan}
\end{figure}

Fig.\ref{Bscan} shows a raster-scanned image of the representative sample using the two orthogonal polarized detectors. Figs.\ref{Bscan}(a) and (b) are the C-scan images of the x- and y-polarized detectors at the time of arrival (TOA) of the reflected pulse from the CFRP substrate, which shows the spatial contrast due to the uneven and polarization-sensitive surface of the substrate. As can be seen, the tiled pattern in Figs.\ref{Bscan}(a) and (b) are complementary, revealing the orientation of the carbon fibers with respect to the polarization of light. Due to the conductivity of carbon, the carbon fibers act as a wire-grid polarizer, and therefore highly reflect the THz beam if they have the same orientation as the polarization of the electric field. Figs.\ref{Bscan}(c) and (d) demonstrate the corresponding B-scans of the solid horizontal lines in the C-scans, formed by the impulse response of the sample. The air-coating interface and the coating-CFRP interface can be observed in the cross-sectional images. The air-coating interface can be seen to have a uniform surface, while the CFRP substrate shows an uneven and wavy surface due to its interwoven nature. Figs.\ref{Bscan}(e) and (f) represent the impulse response B scans along the vertical dashed lines in the subfigures (a) and (b). In the horizontal cross section, the second positive impulse has a spatial wavy pattern, with the waves having alternating high and low reflection. In contrast, on the vertical cross section, this spatial wavy pattern is less observable. This difference in the two vertical and horizontal B-scans, indicating the warp and weft of the interwoven CFRP, suggests the interwoven CFRP is anisotropic. The horizontal cross sections in Fig.\ref{Bscan}(d) and (c) show the wefts of the interwoven structure, while the vertical cross sections in Fig.\ref{Bscan}(e) and (f) demonstrate the warps of the CFRP, which tend to be more stretched \cite{lee2001effect}. Therefore, the "bright tiles" in Fig.\ref{Bscan}(a) indicate the wefts, whereas the bright tiles in the y-polarization channel in Fig.\ref{Bscan}(b) show the warps.

\begin{figure}[!h]
\includegraphics[width=\linewidth]{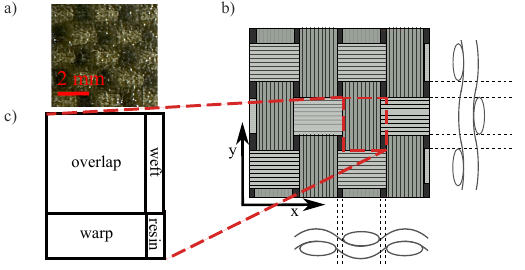}
\caption{Uncoated CFRP interwoven structure. a) Photographic image of the uncoated tiled patterned CFRP. b) Anisotropic interwoven structure. The red dashed box indicates a cell unit. At the bottom and on the right side, cross-sectional structures of the interwoven fibers are shown. c) Segments of a unit cell of interwoven structure.}
\label{CFRPcell}
\end{figure}

This behavior can be better understood using the conceptual drawings in Fig.\ref{CFRPcell}, where the subfigure (a) is a photograph of an uncoated interwoven CFRP showing the tile pattern. Fig.\ref{CFRPcell}(b) shows the anisotropic pattern using the interwoven bundles of the carbon fibers. In particular, the gap between two warp regions can be different from the gap between the weft regions. In other words, if we consider a cell unit as demonstrated by the red dashed box, each cell unit can be segmented into 4 regions as introduced in \cite{mirotznik2011broadband}: warp, weft, resin and overlap of warp and weft, as shown in Fig.\ref{CFRPcell}(c). 

\begin{figure}
\centering
\includegraphics[width=\linewidth]{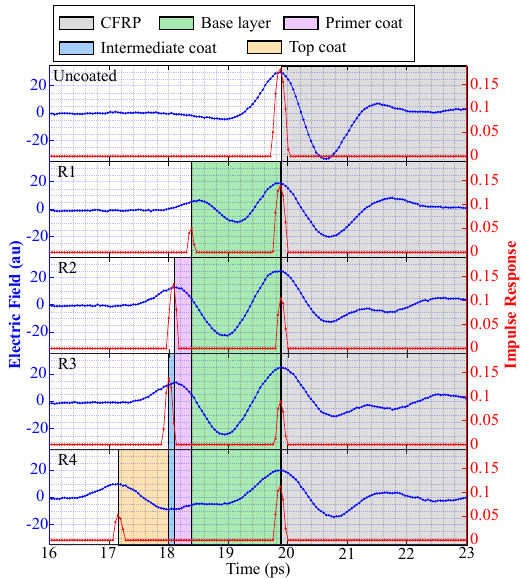}
\caption{THz time-domain pulses (blue) with their corresponding impulse response function (red) over different regions of multilayer step-profile of a sample. From top to bottom, pulses are from the uncoated (backside) region of the sample, R1 (Base Layer), R2 (Primer Coat), R3 (Intermediate Coat), and R4 (Top Coat) regions. The pulses are chosen from the center of the bright tiles to account for non-uniformity and isotropic structure of the sample.}
\label{allpulses}
\end{figure}

To separate the contribution from different coating layers, we used measurements obtained in regions R1 to R4 of the coated samples, as shown by the step profile in Fig.\ref{samplefig}, as well as a measurement from a representative uncoated (bare) CFRP section. Fig.\ref{allpulses} shows the reflected electric field (blue) and the impulse response function (red) of a representative uncoated sample and the R1 to R4 regions, from top to bottom, respectively. The time-domain electric field is chosen from the center of a bright tile measured by the x-polarized detector. The corresponding impulse responses of the electric field measurements were calculated using the ISA deconvolution described in \ref{sparse deconvolution}. As shown, the impulse response of the uncoated sample has only one positive impulse, indicating the air-CFRP interface. 
With the addition of the BL on the top of the carbon-fiber substrate in R1, another impulse appears prior to the CFRP impulse, which accounts for the air-coating interface. The separation of these two impulses (i.e., air-coating and coating-CFRP impulses) increases as more layers are added over the step profile coating. It is worth mentioning that no impulses can be seen between different coating layers, which might be due to the close refractive indices of these layers. The time difference $\Delta t$ between the two impulses will be used for further analysis in Section \ref{prediction of coating thickness}.

\section{Coating Thickness Measurement}
\subsection{Modeling Optical Properties of Coated Layers}
\label{modelingsection}
The interwoven CFRP panels were divided into two groups (training and test); Only 6 training samples were used to model the optical properties of the coating layers and 17 test samples to evaluate the performance of the model. The first author, who conducted the following analysis, was blinded to the actual thickness of the test samples. Polarimetric THz measurements of the four regions of all samples were captured. Only the center pixels of the tiles in a THz scan were used for further processing to ensure that the angle of incidence on the weave pattern was approximately normal. As explained in Section \ref{THz Imaging of CFRP With Multilayered Coating}, the $\Delta t$ between the two positive peaks is extracted from the impulse response function of the center pixels. These $\Delta t$ values can be converted into $OPL$ using (\ref{OPL1eq}), which gives the total $OPL$ of all the coating layers in the corresponding region of the step profile. 
The extracted $OPL$ can be converted to the total thickness of the coating layers, if the index of refraction of each layer is known. In this case, the effective refractive index of all coating layers can be modeled by a weighted average of the sub-layers, given by,
\begin{equation}
\label{avg_n}
n_{avg}=\frac{\sum_{i} n_{i} d_{i}}{\sum_{i} d_{i}},
\end{equation}
where $n_{i}$ and $d_{i}$ are the refractive index and thickness of the $i$th coating layer, respectively. Since no internal reflection echos can be seen in the impulse response function of the sample, shown in Fig.\ref{allpulses}, it is reasonable to assume that the refractive indices of the layers are close to each other. Therefore, we can use a single liner model to relate the extracted $OPL$s from regions R2, R3 and R4 of the training dataset and the actual total thickness of the respective samples, which were provided using the borehole technique by the manufacturer. This linear model, $OPL = n\,d$, is shown in Fig.\ref{fitfig1}(a), where the slope of the line gives the value of $n_{avg}$. Moreover, since the thickness of the BL was not provided (see the step profile in Fig. 2), the non-zero y-intercept of this line corresponds to the OPL of this layer ($204.48\pm14.52 \mu m$). In other words, the parameters of the fitted line can be extracted using the training dataset and the following equation, 
\begin{equation}
\label{fit1eq}
{OPL}_{tot,i}=\hat{n}_{avg}d_{tot,i}+\widehat{OPL}_{BL},
\end{equation}
where $d_{tot,i}$ is the actual total thickness of the $i$th region ($i\in[2, 3, 4])$ from the fitting line, $OPL_{tot,i}$ is the measured total $OPL$, and $\widehat{OPL}_{BL}$ is the modeled $OPL$ of BL as the y-intercept of the equation (note that estimated values are shown with a hat symbol through the paper). Therefore, using the fitting parameters in (\ref{fit1eq}), we can predict the total thickness $\hat{d}_{tot,i}$ in future measurements (i.e., test sample group), 
\begin{equation}
\label{convopl2d}
\hat{d}_{tot,i}=\frac{OPL_{tot,i}-\widehat{OPL}_{BL}}{\hat{n}_{avg}},
\end{equation}
where $OPL_{tot,i}$ is the value extracted from the measurement. This conversion allows us to directly evaluate the performance of the linear model given the ground-truth measurements provided for the test samples. 

In order to model the thickness of each layer separately, we can utilize the step profile of the sample coatings. The $OPL$ of each region 1-4 can be described as,
\begin{equation}
\label{separateOPL1}
\begin{split}
OPL_1&=n_{BL}d_{BL},\\
OPL_2&=n_{BL}d_{BL}+n_{PC}d_{PC},\\
OPL_3&=n_{BL}d_{BL}+n_{PC}d_{PC}+n_{IC}d_{IC},\\
OPL_4&=n_{BL}d_{BL}+n_{PC}d_{PC}+n_{IC}d_{IC}+n_{TC}d_{TC}.
\end{split}
\end{equation}
Subtracting the $OPL$ of each region from the next one, which has one more coating layer, results the $OPL$ of each layer,
\begin{equation}
\label{separateOPL2}
\begin{split}
OPL_{PC}&=OPL_2-OPL_1,\\
OPL_{IC}&=OPL_3-OPL_2,\\
OPL_{TC}&=OPL_4-OPL_3.
\end{split}
\end{equation}
Similarly as before, because the refractive indices of the layers are assumed to be close to each other, a single line can be fitted to $OPL_{PC}$, $OPL_{IC}$, and $OPL_{TC}$ in aggregate. In other words, assuming each layer has the same refractive index $n_{c}$ (c stands for coating), the linear fitting model will be,
\begin{equation}
\label{fitting2}
OPL_{c,i}=\hat{n}_cd_i+\hat{\delta},
\end{equation}
where $OPL_{c,i}$ is the measured OPL of the $i$the coating layer, and $d_i$ is the actual thickness of the layer, and $\hat{\delta}$ is the y-intercept of the fitting model. Although the y-intercept in (\ref{fitting2}) does not represent a physical parameter, it is best not to apply a regression through the origin \cite{eisenhauer2003regression}. Same as (\ref{convopl2d}), we use (\ref{fitting2}) to convert measured $OPL$ of each layer to predicted thickness of the future measurements:
\begin{equation}
\label{convopl2d2}
\hat{d}_i=\frac{OPL_{c,i}-\hat{\delta}}{\hat{n}_c}.
\end{equation}
Based on the assumption of the matching refractive indices of the coating layers, it is expected that the slopes of the two fitted lines in (\ref{fit1eq}) and (\ref{fitting2}), i.e., $\hat{n}_{avg}$ and $\hat{n}_c$, to have relatively the same values.

\subsection{Statistical Metrics}
A first-degree least squares fitting procedure is utilized to model the measured values of the training data set to extract the parameters, described in (\ref{fit1eq}) and (\ref{fitting2}). The coefficient of determination ($R^2$) is calculated to assess how well the models fit the training data. The confidence intervals (CI) of the models are calculated, which reflect the uncertainty of the model parameters ($\hat{n}_{avg}$ and $\widehat{OPL}_{BL}$ in (\ref{fit1eq}), and $\hat{n}_c$ and $\hat{\delta}$ in (\ref{fitting2})) using the training data set. The root of the mean squared error (RMSE) is calculated between the predicted thickness in the test data set and the corresponding actual values. The prediction interval (PI) of the training data set indicates the expected range for future measurements and is applied after the conversion of $OPL$s to the predicted thicknesses. The coverage probability (CP), is a measure of how much of the test data lies within the PI boundaries \cite{khosravi2010lower}, and is calculated by,
\begin{equation}
\label{CPeq}
CP=\frac{1}{N}\sum_ic_i,
\end{equation}
\begin{equation}
\label{ci}
c_i=
\begin{cases}
1 &  d_{act}-PI_L\leq d_{pred}\leq d_{act}+PI_U,\\
0 &  otherwise
\end{cases},
\end{equation}
where $d_{pred}$ and $d_{act}$ are the predicted and actual coating thickness, and $PI_L$ and $PI_U$ are the lower and upper boundaries of PI at the corresponding $d_{act}$, respectively. The mean of prediction interval (MPI) indicates the average distance of PI from the $y=x$ line, and is given by,
\begin{equation}
\label{MPIHW}
MPI=\pm \frac{1}{2N}\sum_i(PI_{U}(x)-PI_{L}(x)).
\end{equation}

In other words, for a 95\% PI, it is expected that the future data set falls within the range of $\pm MPI$ $\mu m$, and the CP assesses how much this expectation is satisfied.

\subsection{Prediction of Coating Thickness}
\label{prediction of coating thickness}
As discussed in Section \ref{THz Imaging of CFRP With Multilayered Coating}, the so-called bright and dark tiles show different reflectivity due to their relative orientation with the polarization of incident light. Therefore, the analysis of bright tiles is conducted separately on the x- and y- polarized channels of the PHASR Scanner to account for the anisotropy of the substrate. Moreover, to explore the effect of the lower signal-to-noise ratio (SNR) over the dark tiles on the accuracy of coating thickness prediction, we separately modeled these pixels using one of the two channels of the scanner. Consequently, there are 3 data sets to analyze: 1) x-polarized detector data over its bright tile regions, 2) x-polarized detector data over its dark tiles, and 3) y-polarized detector data on its bright tiles. 

\begin{figure*}[!t]
\centering
\includegraphics[width=0.8\linewidth]{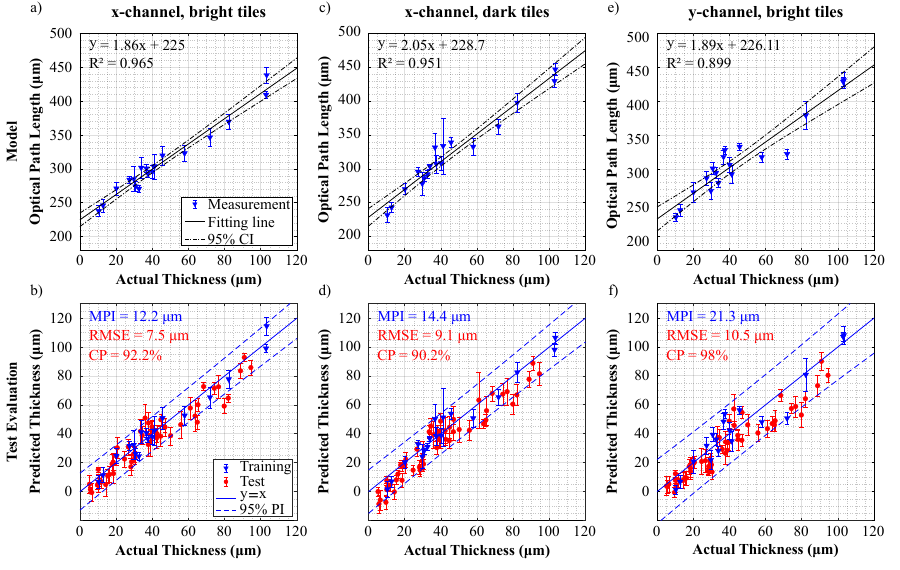}
\caption{Training and evaluation of the model for the total coating thickness measurement of all layers using the measured OPLs. a) Fitting model on the OPL of bright tiles using the x-polarized detector with 95\% CI. b) Training (blue) and test (red) evaluation on the model in sub-figure (a) with 95\% PI. c) Fitting model on the OPL of dark tiles using the x-polarized detector. d) Training (blue) and test (red) evaluation on dark tiles using x-polarized detector. e) Fitting model on the OPL of bright tiles using the y-polarized detector. f) Training and test evaluation on bright tiles from the y-polarized detector.}
\label{fitfig1}
\end{figure*}

Fig.\ref{fitfig1}(a) shows the modeling of the measured $OPL_{tot}$ on the actual total thickness values over bright tiles of the x-polarized detector using (\ref{fit1eq}). As mentioned earlier, the training data used in this figure are only from 6 samples, each contributing 3 thickness values for PC, IC, and TC layers. The fitted line has a slope of $1.86\pm 0.09$ indicating $\hat{n}_{avg}$, and the y-intercept is $225\pm 2.48$ $\mu m$, corresponding to the estimated average $\widehat{OPL}_{BL}$. The error bars show the standard deviation of the measured $OPL$ in the same scan. Fig.\ref{fitfig1}(b) shows the predicted thickness of the training (blue points) and the test data set (red points), after the conversion of $OPL_{tot}$ to $\hat{d}_{tot}$ using (\ref{convopl2d}). The PI is calculated based on $\hat{d}_{tot}$ of the training data set, which has an MPI of $12.2 \mu m$. The CP of the test dataset is 92.2\%, which indicates that 92.2\% of the unseen test data points fall within $\pm12.2 \mu m$ of the predicted line. In addition, the model line has an RMSE of $7.5 \mu m$ using the test data set.

Fig.\ref{fitfig1}(c) shows the fitting model using the training data set over the dark tiles captured by the x-polarized detector. Compared to Fig.\ref{fitfig1}(a), the slope of the fitted line is slightly higher ($2.05\pm 0.12)$, while the y intercept is relatively the same as that of Fig.\ref{fitfig1}(a) ($228.7 \pm 3.21$ $\mu m$). In Fig.\ref{fitfig1}(d), the predicted values of the training and test data sets are plotted compared to the actual thickness values. As it can be seen, the MPI is slightly larger than that of Fig.\ref{fitfig1}(b), and the CP is 90.2\% indicating that the calculated PI from the training data is well represented for the unseen test data set, and future data points should be within the range $\pm 14.4 \mu m$ of the predicted value. The RMSE of the test data set (prediction) with respect to the model is $9.1$ $\mu m$. It can be concluded that, although the SNR is lower over the dark tiles of the sample in contrast to the bright tiles, the performance of the model has not changed significantly compared to Fig.\ref{fitfig1}(b).  

The model trained using the bright tiles from the y-polarized detector data set is shown in Fig.\ref{fitfig1}(e). The $\hat{n}_{avg}$ from the slope of the line, $1.89\pm0.16$, is about the same as the $\hat{n}_{avg}$ from Fig.\ref{fitfig1}(a). Moreover, the y-intercept is reasonably close to that of Fig.\ref{fitfig1}(a) being $226.11 \pm 4.4$ $\mu m$, with $R^2=0.899$. Furthermore, the MPI of the model from the training data set is $\pm21.3 \mu m$, and 98\% of the test data points are within this range. Additionally, the RMSE of the predicted test data set is $10.5 \mu m$, which is close to that of Fig.\ref{fitfig1}(b) and d. In summary, each of these three models could accurately predict the total coating thickness over the CFRP substrate. The linear relationship between $\hat{n}_{avg}$ in (\ref{avg_n}) and $d_{tot}$ of each ROI confirms the hypothesis that we have coating layers with matched refractive indices. Finally, in the worst-case scenario, the CP is better than 90 2\%, and the RMSE is smaller than 10.5 $\mu$m.

\begin{figure*}[!h]
\centering
\includegraphics[width=0.8\linewidth]{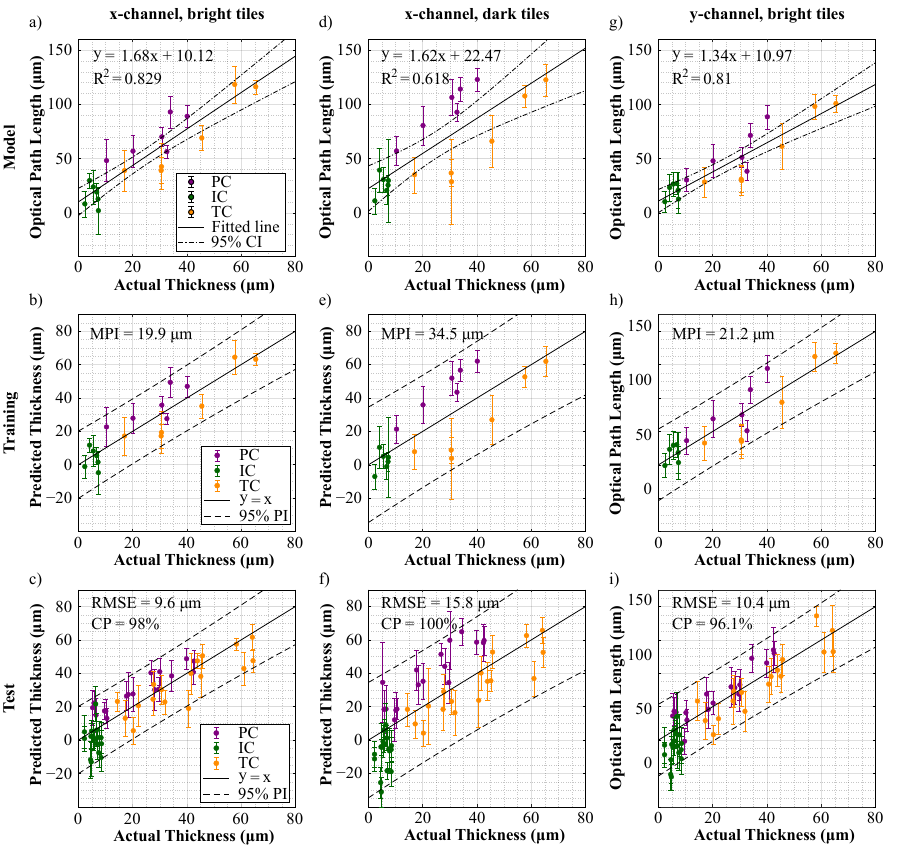}
\caption{Modeling OPL using the training data set and prediction for test data set for individual coating layers. a) Modeling OPL over training data set for each layer of bright tiles using the x-polarized detector. b) Converted OPL into the predicted thickness with 95\% PI using the training data set over the bright tiles of the x-polarized detector. c) Assessment of the predicted thickness using the test data set over bright tiles recorded by the x-polarized detector. d) Modeling OPL using the training data set over the dark tiles recorded by the x-polarized detector. e) Converted OPL into the predicted thickness with 95\% PI using the training data set over the dark tiles of the x-polarized detector. f) Assessment of the predicted thickness using the test data set over the dark tiles recorded by the x-polarized detector. g) Modeling OPL using the training data set over the bright tiles recorded by the y-polarized detector. h) Converted OPL into predicted thickness with 95\% PI using the training data set of the bright tiles of the y-polarized detector . i) Assessment of the predicted thickness using the test data set over the bright tiles recorded by the y-polarized detector.}
\label{fitfig2}
\end{figure*}

Using these three data sets, we can also extract the thickness of the individual layers separately, as described by (\ref{separateOPL1}). Fig.\ref{fitfig2}(a) shows the modeling of the measured $OPL_{c}$ for each coating layer using the training data set of bright tiles from the x-polarized detector. Since the refractive indices of the layers are assumed to be close, a single line was used in this model. The slope and y-intercept of the line indicate $\hat{n}_c=1.68\pm0.19$ and $\hat{\delta}=10.12 \pm 3.65$ according to equation (\ref{fitting2}), respectively, and the data points have an $R^2$ of 0.829. The $OPL$ values are then converted to predicted thickness using (\ref{convopl2d2}), as shown in Fig.\ref{fitfig2}(b). The MPI of the training dataset is $\pm19.9 \mu m$. Fig.\ref{fitfig2}(c) shows the predicted values of each layer in the test data set, showing a coverage probability of 98\% within the $\pm19.9 \mu m$ range, and an RMSE value of $9.6 \mu m$. The 98\% CP indicates an excellent training model based on only a few training samples. Fig.\ref{fitfig2}(d) shows the training model on dark tiles of the x-polarized channel. The slope of the line is $1.62\pm0.32$ with a y-intercept of $22.47\pm6.09$, while the $R^2$ is only 0.618, which is due to the lower SNR of the dark tiles. The training prediction indicates an MPI of $34.5 \mu m$ (Fig.\ref{fitfig2}(e)). The evaluation of the test data set results in 100\% CP and RMSE of 15.8 $\mu$m.

Fig.\ref{fitfig2} (g) shows that the training model of the y-polarized channel on bright tiles results in a lower $\hat{n}_c$ compared to that of Fig.\ref{fitfig2}(a), i.e., $1.34\pm0.16$, and the $R^2$ of 0.81. The MPI of the predicted thickness from the training data set is $\pm21.2 \mu m$, which is close to that of the other detector (Fig.\ref{fitfig2}(h)). The analysis of the test data set shows an RMSE of $10.4 \mu m$ and a CP of 96. 1\%. The $R^2$ values obtained in Fig.\ref{fitfig2} are generally smaller than those in Fig.\ref{fitfig1}. In addition, other metrics including the size of error bars, PI and RMSE are better in Fig.\ref{fitfig1} compared to Fig.\ref{fitfig2}. This might be due to the propagation of error, which is made by subtracting the measured values from each other, and indicating the limitation of the technique in measuring smaller thicknesses.

\section{Discussion}
This study explores the potential use of a THz polarization-sensitive PHASR Scanner in understanding the structure of an interwoven CFRP substrate coated with multiple layers. The use of two polarization data channels in tandem with the ISA sparse deconvolution showed the anisotropic response of the fiber bundles of the CFRP, and revealed the large effect of the weave pattern on extraction of the accurate coating thickness. Moreover, the linear training model demonstrated the possibility of extracting not only the thickness of the coating layers with an accuracy of more than 90\% within a minimum margin of about $\pm12 \mu m$, but also estimating the average refractive index of the coating layers. We hypothesized that the PC, IC and TC layers have matched refractive indices because the expected multiple internal reflection echos were not evident in the impulse response function of the sample. The overall training models (total coating thickness) in Fig.\ref{fitfig1} show that although the slopes of the lines vary slightly between the three models, the extracted $n_{ave}$ values are fairly close considering the confidence interval of the statistical model. Additionally, the performance metrics from the x- and y-polarized detectors are slightly different, which might be due to anisotropic structure of the sample and the effect of the weave pattern. This difference in the accuracy of the models suggests that the orientation of the sample with respect to the scanner should be carefully considered.

The use of ISA sparse deconvolution provides a clear picture of the boundary reflections within the multilayered structures. However, as was seen in this study, having multiple coatings of the same optical properties can result in no impulses between layers. In addition, the dispersion of the sample, which results in a widening of the impulse response \cite{szabo2003material}, cannot be easily taken into account due to the sparsity constraint. This pulse widening can be modeled in ISA sparse deconvolution by defining a widened impulse function kernel \cite{dong2017terahertzdeconv, olofsson2001minimum}. However, not only does this method require presumption of a kernel impulse for the sample, but also requires self-deconvolution of the kernel impulse function, which increases the computation time. Furthermore, the complex refractive index of a medium causes a dispersive phase shift in the transfer function of the sample \cite{bernier2018comparative}, resulting in a distorted impulse response. 
Finally, one key limitation of any broad time-domain technique would be the bandwidth of the emitter and detectors used in the scanner, which in this study was limited to about 1.5 THz. Expanding the bandwidth of THz polarimetry systems, for example, using plasma-based sources and EO-sampling crystals \cite{xu_broadband_2022}, can further reduce this resolution limit.

\section{Conclusion}
We presented polarization-sensitive THz time-domain spectroscopic imaging of carbon-fiber-reinforced polymer samples with multiple coating layers. The time-of-flight analysis of the THz images in each polarization channel, which was extracted by utilizing the sparse deconvolution of the THz-TDS pulses, in parallel with the polarimetric spatial information of the substrate, provided a clear insight to the nonuniform and anisotropic response of the sample. Furthermore, a simple physics-based linear regression model was introduced to accurately extract the thickness of the coating layers, in addition to the estimation of their optical properties. We showed that training the model parameters using only 6 samples was sufficient to achieve a coverage probability better than 90.2\% and RMSE smaller than 10.5 $\mu$m.

\bibliography{coating_thickness}

\begin{thebibliography}{10}
\providecommand{\url}[1]{#1}
\csname url@samestyle\endcsname
\providecommand{\newblock}{\relax}
\providecommand{\bibinfo}[2]{#2}
\providecommand{\BIBentrySTDinterwordspacing}{\spaceskip=0pt\relax}
\providecommand{\BIBentryALTinterwordstretchfactor}{4}
\providecommand{\BIBentryALTinterwordspacing}{\spaceskip=\fontdimen2\font plus
\BIBentryALTinterwordstretchfactor\fontdimen3\font minus \fontdimen4\font\relax}
\providecommand{\BIBforeignlanguage}[2]{{%
\expandafter\ifx\csname l@#1\endcsname\relax
\typeout{** WARNING: IEEEtran.bst: No hyphenation pattern has been}%
\typeout{** loaded for the language `#1'. Using the pattern for}%
\typeout{** the default language instead.}%
\else
\language=\csname l@#1\endcsname
\fi
#2}}
\providecommand{\BIBdecl}{\relax}
\BIBdecl

\bibitem{hintzsche2013nonionizing}
H.~Hintzsche, C.~Jastrow, B.~Heinen, K.~Baaske, T.~Kleine-Ostmann, M.~Schwerdtfeger, M.~K. Shakfa, U.~K{\"a}rst, M.~Koch, T.~Schrader \emph{et~al.}, ``Terahertz radiation at 0.380 thz and 2.520 thz does not lead to dna damage in skin cells in vitro,'' \emph{Radiation Research}, vol. 179, no.~1, pp. 38--45, 2013.

\bibitem{tao2020THz}
Y.~H. Tao, A.~J. Fitzgerald, and V.~P. Wallace, ``Non-contact, non-destructive testing in various industrial sectors with terahertz technology,'' \emph{Sensors}, vol.~20, no.~3, p. 712, 2020.

\bibitem{sitnikov2021effects}
D.~S. Sitnikov, I.~V. Ilina, V.~A. Revkova, S.~A. Rodionov, S.~A. Gurova, R.~O. Shatalova, A.~V. Kovalev, A.~V. Ovchinnikov, O.~V. Chefonov, M.~A. Konoplyannikov \emph{et~al.}, ``Effects of high intensity non-ionizing terahertz radiation on human skin fibroblasts,'' \emph{Biomedical optics express}, vol.~12, no.~11, pp. 7122--7138, 2021.

\bibitem{jepsen2011terahertzreview}
P.~U. Jepsen, D.~G. Cooke, and M.~Koch, ``Terahertz spectroscopy and imaging--modern techniques and applications,'' \emph{Laser \& Photonics Reviews}, vol.~5, no.~1, pp. 124--166, 2011.

\bibitem{khani2023burn}
M.~E. Khani, Z.~B. Harris, O.~B. Osman, A.~J. Singer, and M.~Hassan~Arbab, ``Triage of in vivo burn injuries and prediction of wound healing outcome using neural networks and modeling of the terahertz permittivity based on the double debye dielectric parameters,'' \emph{Biomedical Optics Express}, vol.~14, no.~2, pp. 918--931, 2023.

\bibitem{chen2022cornea}
A.~Chen, Z.~B. Harris, A.~Virk, A.~Abazari, K.~Varadaraj, R.~Honkanen, and M.~H. Arbab, ``Assessing corneal endothelial damage using terahertz time-domain spectroscopy and support vector machines,'' \emph{Sensors}, vol.~22, no.~23, p. 9071, 2022.

\bibitem{cheon2017cancer}
H.~Cheon, H.-J. Yang, and J.-H. Son, ``Toward clinical cancer imaging using terahertz spectroscopy,'' \emph{IEEE Journal of Selected Topics in Quantum Electronics}, vol.~23, no.~4, pp. 1--9, 2017.

\bibitem{virk2021development}
A.~S. Virk, Z.~B. Harris, and M.~H. Arbab, ``Development of a terahertz time-domain scanner for topographic imaging of spherical targets,'' \emph{Optics letters}, vol.~46, no.~5, pp. 1065--1068, 2021.

\bibitem{bowman2016terahertz}
T.~Bowman, M.~El-Shenawee, and L.~K. Campbell, ``Terahertz transmission vs reflection imaging and model-based characterization for excised breast carcinomas,'' \emph{Biomedical Optics Express}, vol.~7, no.~9, pp. 3756--3783, 2016.

\bibitem{tzydynzhapov2020security}
G.~Tzydynzhapov, P.~Gusikhin, V.~Muravev, A.~Dremin, Y.~Nefyodov, and I.~Kukushkin, ``New real-time sub-terahertz security body scanner,'' \emph{Journal of Infrared, Millimeter, and Terahertz Waves}, vol.~41, pp. 632--641, 2020.

\bibitem{li2018study}
R.~Li, C.~Li, H.~Li, S.~Wu, and G.~Fang, ``Study of automatic detection of concealed targets in passive terahertz images for intelligent security screening,'' \emph{IEEE Transactions on Terahertz Science and Technology}, vol.~9, no.~2, pp. 165--176, 2018.

\bibitem{cheng2022improved}
L.~Cheng, Y.~Ji, C.~Li, X.~Liu, and G.~Fang, ``Improved ssd network for fast concealed object detection and recognition in passive terahertz security images,'' \emph{Scientific Reports}, vol.~12, no.~1, p. 12082, 2022.

\bibitem{shchepetilnikov2020new}
A.~Shchepetilnikov, P.~Gusikhin, V.~Muravev, G.~Tsydynzhapov, Y.~A. Nefyodov, A.~Dremin, and I.~Kukushkin, ``New ultra-fast sub-terahertz linear scanner for postal security screening,'' \emph{Journal of Infrared, Millimeter, and Terahertz Waves}, vol.~41, pp. 655--664, 2020.

\bibitem{amenabar2013introductory}
I.~Amenabar, F.~Lopez, and A.~Mendikute, ``In introductory review to thz non-destructive testing of composite mater,'' \emph{Journal of Infrared, Millimeter, and Terahertz Waves}, vol.~34, pp. 152--169, 2013.

\bibitem{yakovlev2015non}
E.~V. Yakovlev, K.~I. Zaytsev, I.~N. Dolganova, and S.~O. Yurchenko, ``Non-destructive evaluation of polymer composite materials at the manufacturing stage using terahertz pulsed spectroscopy,'' \emph{IEEE Transactions on Terahertz science and Technology}, vol.~5, no.~5, pp. 810--816, 2015.

\bibitem{palka2016non}
N.~Palka, R.~Panowicz, M.~Chalimoniuk, and R.~Beigang, ``Non-destructive evaluation of puncture region in polyethylene composite by terahertz and x-ray radiation,'' \emph{Composites Part B: Engineering}, vol.~92, pp. 315--325, 2016.

\bibitem{wang2022real}
X.~Wang, Z.~Zhang, Y.~Xu, L.~Zhang, R.~Yan, and X.~Chen, ``Real-time terahertz characterization of minor defects by the yolox-msa network,'' \emph{IEEE Transactions on Instrumentation and Measurement}, vol.~71, pp. 1--10, 2022.

\bibitem{mei2021terahertz}
H.~Mei, H.~Jiang, F.~Yin, L.~Wang, and M.~Farzaneh, ``Terahertz imaging method for composite insulator defects based on edge detection algorithm,'' \emph{IEEE Transactions on Instrumentation and Measurement}, vol.~70, pp. 1--10, 2021.

\bibitem{dong2015nondestructive}
J.~Dong, B.~Kim, A.~Locquet, P.~McKeon, N.~Declercq, and D.~Citrin, ``Nondestructive evaluation of forced delamination in glass fiber-reinforced composites by terahertz and ultrasonic waves,'' \emph{Composites Part B: Engineering}, vol.~79, pp. 667--675, 2015.

\bibitem{cheng2022contamination}
H.~Cheng, H.-c. Huang, M.-f. Yang, M.-h. Yang, H.~Yan, S.~Panezai, Z.-Y. Zheng, Z.~Zhang, and Z.-l. Zhang, ``Characterization of the remediation of chromium ion contamination with bentonite by terahertz time-domain spectroscopy,'' \emph{Scientific Reports}, vol.~12, no.~1, p. 11149, 2022.

\bibitem{lu2022contamination}
W.~Lu, H.~Luo, L.~He, W.~Duan, Y.~Tao, X.~Wang, and S.~Li, ``Detection of heavy metals in vegetable soil based on thz spectroscopy,'' \emph{Computers and Electronics in Agriculture}, vol. 197, p. 106923, 2022.

\bibitem{stelmaszczyk2022ultrafast}
K.~Stelmaszczyk, E.~Karpierz-Marczewska, V.~Mikhnev, G.~Cywinski, T.~Skotnicki, and W.~Knap, ``Ultrafast time-of-flight method of gasoline contamination detection down to ppm levels by means of terahertz time-domain spectroscopy,'' \emph{Applied Sciences}, vol.~12, no.~3, p. 1629, 2022.

\bibitem{khani2021diffuse}
M.~E. Khani, O.~B. Osman, and M.~H. Arbab, ``Diffuse terahertz spectroscopy in turbid media using a wavelet-based bimodality spectral analysis,'' \emph{Scientific Reports}, vol.~11, no.~1, p. 22804, 2021.

\bibitem{sun2020exploiting}
Q.~Sun, X.~Chen, X.~Liu, R.~I. Stantchev, and E.~Pickwell-MacPherson, ``Exploiting total internal reflection geometry for terahertz devices and enhanced sample characterization,'' \emph{Advanced Optical Materials}, vol.~8, no.~3, p. 1900535, 2020.

\bibitem{palka2016characterization}
N.~Palka, M.~Szala, and E.~Czerwinska, ``Characterization of prospective explosive materials using terahertz time-domain spectroscopy,'' \emph{Applied optics}, vol.~55, no.~17, pp. 4575--4583, 2016.

\bibitem{khani2021chemical}
M.~E. Khani and M.~H. Arbab, ``Chemical identification in the specular and off-specular rough-surface scattered terahertz spectra using wavelet shrinkage,'' \emph{IEEE Access}, vol.~9, pp. 29\,746--29\,754, 2021.

\bibitem{khani2022multiresolution}
M.~E. Khani, Z.~B. Harris, M.~Liu, and M.~H. Arbab, ``Multiresolution spectrally-encoded terahertz reflection imaging through a highly diffusive cloak,'' \emph{Optics Express}, vol.~30, no.~18, pp. 31\,550--31\,566, 2022.

\bibitem{zhou2021effective}
J.~W. Zhou and M.~H. Arbab, ``Effective debye relaxation models for binary solutions of polar liquids at terahertz frequencies,'' \emph{Physical Chemistry Chemical Physics}, vol.~23, no.~7, pp. 4426--4436, 2021.

\bibitem{zhong2019progress}
S.~Zhong, ``Progress in terahertz nondestructive testing: A review,'' \emph{Frontiers of Mechanical Engineering}, vol.~14, no.~3, pp. 273--281, 2019.

\bibitem{zhai2020nondestructive}
M.~Zhai, A.~Locquet, C.~Roquelet, P.~Alexandre, L.~Daheron, and D.~Citrin, ``Nondestructive measurement of mill-scale thickness on steel by terahertz time-of-flight tomography,'' \emph{Surface and Coatings Technology}, vol. 393, p. 125765, 2020.

\bibitem{lin2017measurement}
H.~Lin, Y.~Dong, D.~Markl, B.~M. Williams, Y.~Zheng, Y.~Shen, and J.~A. Zeitler, ``Measurement of the intertablet coating uniformity of a pharmaceutical pan coating process with combined terahertz and optical coherence tomography in-line sensing,'' \emph{Journal of Pharmaceutical Sciences}, vol. 106, no.~4, pp. 1075--1084, 2017.

\bibitem{burford2014terahertz}
N.~M. Burford, M.~O. El-Shenawee, C.~B. O’neal, and K.~J. Olejniczak, ``Terahertz imaging for nondestructive evaluation of packaged power electronic devices,'' \emph{Int. J. Emerg. Technol. Adv. Eng}, vol.~4, no.~1, pp. 395--401, 2014.

\bibitem{true2021terahertz}
J.~True, C.~Xi, N.~Jessurun, K.~Ahi, M.~Tehranipoor, and N.~Asadizanjani, ``Terahertz based machine learning approach to integrated circuit assurance,'' in \emph{2021 IEEE 71st Electronic Components and Technology Conference (ECTC)}.\hskip 1em plus 0.5em minus 0.4em\relax IEEE, 2021, pp. 2235--2245.

\bibitem{shur2019sub}
M.~Shur, S.~Rudin, G.~Rupper, M.~Reed, and J.~Suarez, ``Sub-terahertz testing of millimeter wave monolithic and very large scale integrated circuits,'' \emph{Solid-State Electronics}, vol. 155, pp. 44--48, 2019.

\bibitem{zhong2011non}
S.~Zhong, Y.-C. Shen, L.~Ho, R.~K. May, J.~A. Zeitler, M.~Evans, P.~F. Taday, M.~Pepper, T.~Rades, K.~C. Gordon \emph{et~al.}, ``Non-destructive quantification of pharmaceutical tablet coatings using terahertz pulsed imaging and optical coherence tomography,'' \emph{Optics and Lasers in Engineering}, vol.~49, no.~3, pp. 361--365, 2011.

\bibitem{kawase2013non}
M.~Kawase, K.~Yamamoto, K.~Takagi, R.~Yasuda, M.~Ogawa, Y.~Hatsuda, S.~Kawanishi, Y.~Hirotani, M.~Myotoku, Y.~Urashima \emph{et~al.}, ``Non-destructive evaluation method of pharmaceutical tablet by terahertz-time-domain spectroscopy: application to sound-alike medicines,'' \emph{Journal of Infrared, Millimeter, and Terahertz Waves}, vol.~34, pp. 566--571, 2013.

\bibitem{moradikouchi2022terahertz}
A.~Moradikouchi, A.~Spar{\'e}n, S.~Folestad, J.~Stake, and H.~Rodilla, ``Terahertz frequency domain sensing for fast porosity measurement of pharmaceutical tablets,'' \emph{International Journal of Pharmaceutics}, vol. 618, p. 121579, 2022.

\bibitem{krimi2016highly}
S.~Krimi, J.~Klier, J.~Jonuscheit, G.~Von~Freymann, R.~Urbansky, and R.~Beigang, ``Highly accurate thickness measurement of multi-layered automotive paints using terahertz technology,'' \emph{Applied Physics Letters}, vol. 109, no.~2, 2016.

\bibitem{ketelsen2022thz}
H.~Ketelsen, R.~M{\"a}stle, L.~Liebermeister, R.~Kohlhaas, and B.~Globisch, ``Thz time-domain ellipsometer for material characterization and paint quality control with more than 5 thz bandwidth,'' \emph{Applied Sciences}, vol.~12, no.~8, p. 3744, 2022.

\bibitem{dong2015non}
Y.~Dong, J.~Zhang, Y.-c. Shen, K.~Su, and J.~A. Zeitler, ``Non-destructive characterization of automobile car paints using terahertz pulsed imaging and infrared optical coherence tomography,'' in \emph{2015 40th International Conference on Infrared, Millimeter, and Terahertz waves (IRMMW-THz)}.\hskip 1em plus 0.5em minus 0.4em\relax IEEE, 2015, pp. 1--2.

\bibitem{ospald2014aeronautics}
F.~Ospald, W.~Zouaghi, R.~Beigang, C.~Matheis, J.~Jonuscheit, B.~Recur, J.-P. Guillet, P.~Mounaix, W.~Vleugels, P.~V. Bosom \emph{et~al.}, ``Aeronautics composite material inspection with a terahertz time-domain spectroscopy system,'' \emph{Optical Engineering}, vol.~53, no.~3, pp. 031\,208--031\,208, 2014.

\bibitem{cristofani2014nondestructive}
E.~Cristofani, F.~Friederich, S.~Wohnsiedler, C.~Matheis, J.~Jonuscheit, M.~Vandewal, and R.~Beigang, ``Nondestructive testing potential evaluation of a terahertz frequency-modulated continuous-wave imager for composite materials inspection,'' \emph{Optical Engineering}, vol.~53, no.~3, pp. 031\,211--031\,211, 2014.

\bibitem{wang2019nondestructive}
Q.~Wang, X.~Li, T.~Chang, J.~Zhang, L.~Liu, H.~Zhou, and J.~Bai, ``Nondestructive imaging of hidden defects in aircraft sandwich composites using terahertz time-domain spectroscopy,'' \emph{Infrared Physics \& Technology}, vol.~97, pp. 326--340, 2019.

\bibitem{ryu2016nondestructive}
C.-H. Ryu, S.-H. Park, D.-H. Kim, K.-Y. Jhang, and H.-S. Kim, ``Nondestructive evaluation of hidden multi-delamination in a glass-fiber-reinforced plastic composite using terahertz spectroscopy,'' \emph{Composite Structures}, vol. 156, pp. 338--347, 2016.

\bibitem{yasui2005terahertz}
T.~Yasui, T.~Yasuda, K.-i. Sawanaka, and T.~Araki, ``Terahertz paintmeter for noncontact monitoring of thickness and drying progress in paint film,'' \emph{Applied Optics}, vol.~44, no.~32, pp. 6849--6856, 2005.

\bibitem{garcia2011non}
J.~Garc{\'\i}a-Mart{\'\i}n, J.~G{\'o}mez-Gil, and E.~V{\'a}zquez-S{\'a}nchez, ``Non-destructive techniques based on eddy current testing,'' \emph{Sensors}, vol.~11, no.~3, pp. 2525--2565, 2011.

\bibitem{thompson2016x}
A.~Thompson, I.~Maskery, and R.~K. Leach, ``X-ray computed tomography for additive manufacturing: a review,'' \emph{Measurement Science and Technology}, vol.~27, no.~7, p. 072001, 2016.

\bibitem{naresh2020use}
K.~Naresh, K.~Khan, R.~Umer, and W.~J. Cantwell, ``The use of x-ray computed tomography for design and process modeling of aerospace composites: A review,'' \emph{Materials \& Design}, vol. 190, p. 108553, 2020.

\bibitem{dong2018visualization}
J.~Dong, P.~Pomar{\`e}de, L.~Chehami, A.~Locquet, F.~Meraghni, N.~F. Declercq, and D.~Citrin, ``Visualization of subsurface damage in woven carbon fiber-reinforced composites using polarization-sensitive terahertz imaging,'' \emph{NDT \& E International}, vol.~99, pp. 72--79, 2018.

\bibitem{su2014terahertz}
K.~Su, Y.-C. Shen, and J.~A. Zeitler, ``Terahertz sensor for non-contact thickness and quality measurement of automobile paints of varying complexity,'' \emph{IEEE transactions on terahertz science and technology}, vol.~4, no.~4, pp. 432--439, 2014.

\bibitem{harris2024handheld}
Z.~B. Harris, K.~Xu, and M.~H. Arbab, ``A handheld polarimetric imaging device and calibration technique for accurate mapping of terahertz stokes vectors,'' \emph{Scientific Reports}, vol.~14, no.~1, p. 17714, 2024.

\bibitem{harris2022terahertz}
Z.~B. Harris and M.~H. Arbab, ``Terahertz phasr scanner with 2 khz, 100 ps time-domain trace acquisition rate and an extended field-of-view based on a heliostat design,'' \emph{IEEE transactions on terahertz science and technology}, vol.~12, no.~6, pp. 619--632, 2022.

\bibitem{khani2022accurate}
M.~E. Khani, O.~B. Osman, Z.~B. Harris, A.~Chen, J.~W. Zhou, A.~J. Singer, and M.~H. Arbab, ``Accurate and early prediction of the wound healing outcome of burn injuries using the wavelet shannon entropy of terahertz time-domain waveforms,'' \emph{Journal of Biomedical Optics}, vol.~27, no.~11, pp. 116\,001--116\,001, 2022.

\bibitem{khani2022supervised}
M.~E. Khani, Z.~B. Harris, O.~B. Osman, J.~W. Zhou, A.~Chen, A.~J. Singer, and M.~H. Arbab, ``Supervised machine learning for automatic classification of in vivo scald and contact burn injuries using the terahertz portable handheld spectral reflection (phasr) scanner,'' \emph{Scientific Reports}, vol.~12, no.~1, p. 5096, 2022.

\bibitem{osman2022deep}
O.~B. Osman, Z.~B. Harris, M.~E. Khani, J.~W. Zhou, A.~Chen, A.~J. Singer, and M.~Hassan~Arbab, ``Deep neural network classification of in vivo burn injuries with different etiologies using terahertz time-domain spectral imaging,'' \emph{Biomedical Optics Express}, vol.~13, no.~4, pp. 1855--1868, 2022.

\bibitem{virk2023design}
A.~S. Virk, Z.~B. Harris, and M.~H. Arbab, ``Design and characterization of a hyperbolic-elliptical lens pair in a rapid beam steering system for single-pixel terahertz spectral imaging of the cornea,'' \emph{Optics Express}, vol.~31, no.~24, pp. 39\,568--39\,582, 2023.

\bibitem{xu2024thz}
K.~Xu, Z.~B. Harris, P.~Vahey, and M.~H. Arbab, ``Thz polarimetric imaging of carbon fiber-reinforced composites using the portable handled spectral reflection (phasr) scanner,'' \emph{Sensors}, vol.~24, no.~23, p. 7467, 2024.

\bibitem{cao2019noncontact}
B.~Cao, M.~Wang, X.~Li, M.~Fan, and G.~Tian, ``Noncontact thickness measurement of multilayer coatings on metallic substrate using pulsed terahertz technology,'' \emph{IEEE Sensors Journal}, vol.~20, no.~6, pp. 3162--3171, 2019.

\bibitem{scheller2009analyzing}
M.~Scheller, C.~Jansen, and M.~Koch, ``Analyzing sub-100-$\mu$m samples with transmission terahertz time domain spectroscopy,'' \emph{Optics Communications}, vol. 282, no.~7, pp. 1304--1306, 2009.

\bibitem{bourguignon2011sparse}
S.~Bourguignon, C.~Soussen, H.~Carfantan, and J.~Idier, ``Sparse deconvolution: Comparison of statistical and deterministic approaches,'' in \emph{2011 IEEE Statistical Signal Processing Workshop (SSP)}.\hskip 1em plus 0.5em minus 0.4em\relax IEEE, 2011, pp. 317--320.

\bibitem{beck2009ITS}
A.~Beck and M.~Teboulle, ``A fast iterative shrinkage-thresholding algorithm for linear inverse problems,'' \emph{SIAM journal on imaging sciences}, vol.~2, no.~1, pp. 183--202, 2009.

\bibitem{natarajan1995sparse}
B.~K. Natarajan, ``Sparse approximate solutions to linear systems,'' \emph{SIAM journal on computing}, vol.~24, no.~2, pp. 227--234, 1995.

\bibitem{taylor1979deconvolution}
H.~L. Taylor, S.~C. Banks, and J.~F. McCoy, ``Deconvolution with the $\ell$-1 norm,'' \emph{Geophysics}, vol.~44, no.~1, pp. 39--52, 1979.

\bibitem{zuo2013sparsedeconv}
W.~Zuo, D.~Meng, L.~Zhang, X.~Feng, and D.~Zhang, ``A generalized iterated shrinkage algorithm for non-convex sparse coding,'' in \emph{Proceedings of the IEEE international conference on computer vision}, 2013, pp. 217--224.

\bibitem{candes2008deconvolution}
E.~J. Candes, M.~B. Wakin, and S.~P. Boyd, ``Enhancing sparsity by reweighted $\ell -1$ minimization,'' \emph{Journal of Fourier analysis and applications}, vol.~14, pp. 877--905, 2008.

\bibitem{yarlagadda1985fast}
R.~Yarlagadda, J.~Bednar, and T.~Watt, ``Fast algorithms for $\ell$-p deconvolution,'' \emph{IEEE transactions on acoustics, speech, and signal processing}, vol.~33, no.~1, pp. 174--182, 1985.

\bibitem{dong2017terahertzdeconv}
J.~Dong, X.~Wu, A.~Locquet, and D.~S. Citrin, ``Terahertz superresolution stratigraphic characterization of multilayered structures using sparse deconvolution,'' \emph{IEEE Transactions on Terahertz Science and Technology}, vol.~7, no.~3, pp. 260--267, 2017.

\bibitem{wei2009sparseotherapps}
L.~Wei, Z.-y. Huang, and P.-w. Que, ``Sparse deconvolution method for improving the time-resolution of ultrasonic nde signals,'' \emph{Ndt \& E International}, vol.~42, no.~5, pp. 430--434, 2009.

\bibitem{hugelier2016sparse}
S.~Hugelier, J.~J. De~Rooi, R.~Bernex, S.~Duw{\'e}, O.~Devos, M.~Sliwa, P.~Dedecker, P.~H. Eilers, and C.~Ruckebusch, ``Sparse deconvolution of high-density super-resolution images,'' \emph{Scientific reports}, vol.~6, no.~1, p. 21413, 2016.

\bibitem{bronstein2005blind}
M.~M. Bronstein, A.~M. Bronstein, M.~Zibulevsky, and Y.~Y. Zeevi, ``Blind deconvolution of images using optimal sparse representations,'' \emph{IEEE Transactions on Image Processing}, vol.~14, no.~6, pp. 726--736, 2005.

\bibitem{dietz2014all}
R.~J. Dietz, N.~Vieweg, T.~Puppe, A.~Zach, B.~Globisch, T.~G{\"o}bel, P.~Leisching, and M.~Schell, ``All fiber-coupled thz-tds system with khz measurement rate based on electronically controlled optical sampling,'' \emph{Optics letters}, vol.~39, no.~22, pp. 6482--6485, 2014.

\bibitem{lee2001effect}
B.~Lee, K.~Leong, and I.~Herszberg, ``Effect of weaving on the tensile properties of carbon fibre tows and woven composites,'' \emph{Journal of reinforced plastics and composites}, vol.~20, no.~8, pp. 652--670, 2001.

\bibitem{mirotznik2011broadband}
M.~S. Mirotznik, S.~Yarlagadda, R.~McCauley, and P.~Pa, ``Broadband electromagnetic modeling of woven fabric composites,'' \emph{IEEE Transactions on Microwave Theory and Techniques}, vol.~60, no.~1, pp. 158--169, 2011.

\bibitem{eisenhauer2003regression}
J.~G. Eisenhauer, ``Regression through the origin,'' \emph{Teaching statistics}, vol.~25, no.~3, pp. 76--80, 2003.

\bibitem{khosravi2010lower}
A.~Khosravi, S.~Nahavandi, D.~Creighton, and A.~F. Atiya, ``Lower upper bound estimation method for construction of neural network-based prediction intervals,'' \emph{IEEE transactions on neural networks}, vol.~22, no.~3, pp. 337--346, 2010.

\bibitem{szabo2003material}
T.~L. Szabo, ``The material impulse response for broadband pulses in lossy media,'' in \emph{IEEE Symposium on Ultrasonics, 2003}, vol.~1.\hskip 1em plus 0.5em minus 0.4em\relax IEEE, 2003, pp. 748--751.

\bibitem{olofsson2001minimum}
T.~Olofsson and T.~Stepinski, ``Minimum entropy deconvolution of pulse-echo signals acquired from attenuative layered media,'' \emph{The Journal of the Acoustical Society of America}, vol. 109, no.~6, pp. 2831--2839, 2001.

\bibitem{bernier2018comparative}
M.~Bernier, F.~Garet, E.~Kato, B.~Blampey, and J.-L. Coutaz, ``Comparative study of material parameter extraction using terahertz time-domain spectroscopy in transmission and in reflection,'' \emph{Journal of Infrared, Millimeter, and Terahertz Waves}, vol.~39, pp. 349--366, 2018.

\bibitem{xu_broadband_2022}
K.~Xu, M.~Liu, and M.~H. Arbab, ``Broadband terahertz time-domain polarimetry based on air plasma filament emissions and spinning electro-optic sampling in gap,'' \emph{Applied Physics Letters}, vol. 120, no.~18, 2022.

\end{thebibliography}
\bibliographystyle{IEEEtran}
\begin{IEEEbiography}
[{\includegraphics[width=1in,height=1.25in,clip,keepaspectratio]{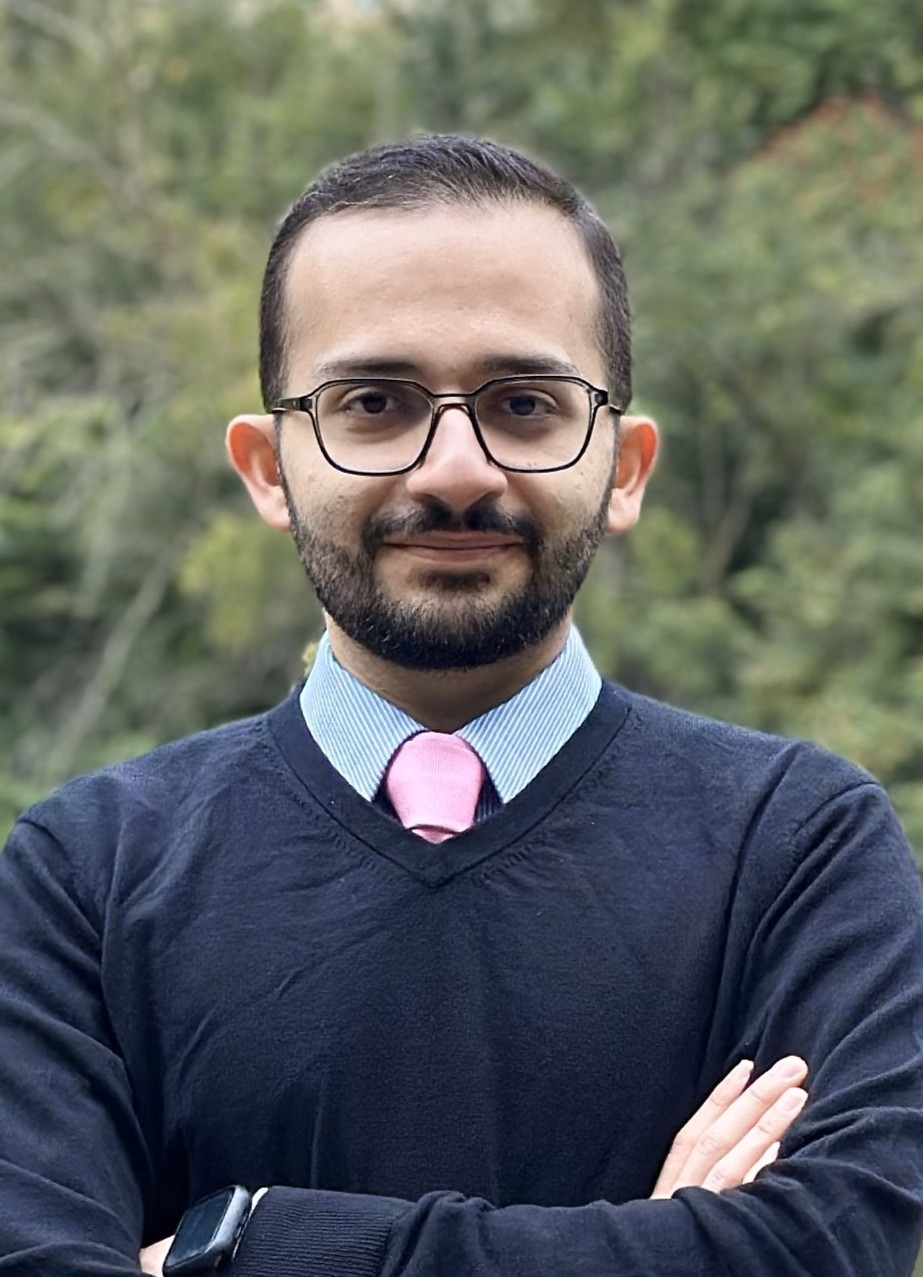}}]{Arash Karimi} received the B.S. degree in electrical engineering from the University of Tehran, Tehran, Iran, in 2022. 

He is currently pursuing a Ph.D. degree in biomedical engineering at Stony Brook University, Stony Brook, NY, USA. His research interests include terahertz spectroscopy and imaging, signal processing, machine learning, non-destructive testing, and biomedical imaging.
\end{IEEEbiography}

\begin{IEEEbiography}
[{\includegraphics[width=1in,height=1.25in,clip,keepaspectratio]{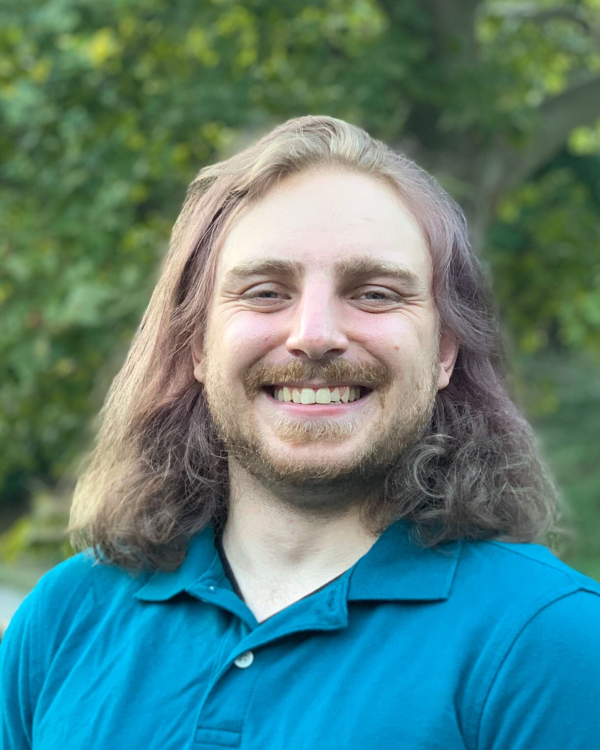}}]{Zachery B. Harris} received the B.S. degree in physics from the University of Washington, Seattle, WA, USA, in 2017. 
From 2018 to 2023 he was a Research Technician in the department of Biomedical Engineering at Stony Brook University, Stony Brook, New York, USA. 

He is currently pursuing a Ph.D. degree in biomedical engineering at Stony Brook University, Stony Brook, NY, USA. His research interests include terahertz imaging and spectroscopy, terahertz signal processing, non-destructive testing and evaluation, and medical imaging.
\end{IEEEbiography}

\begin{IEEEbiography}
[{\includegraphics[width=1in,height=1.25in,clip,keepaspectratio]{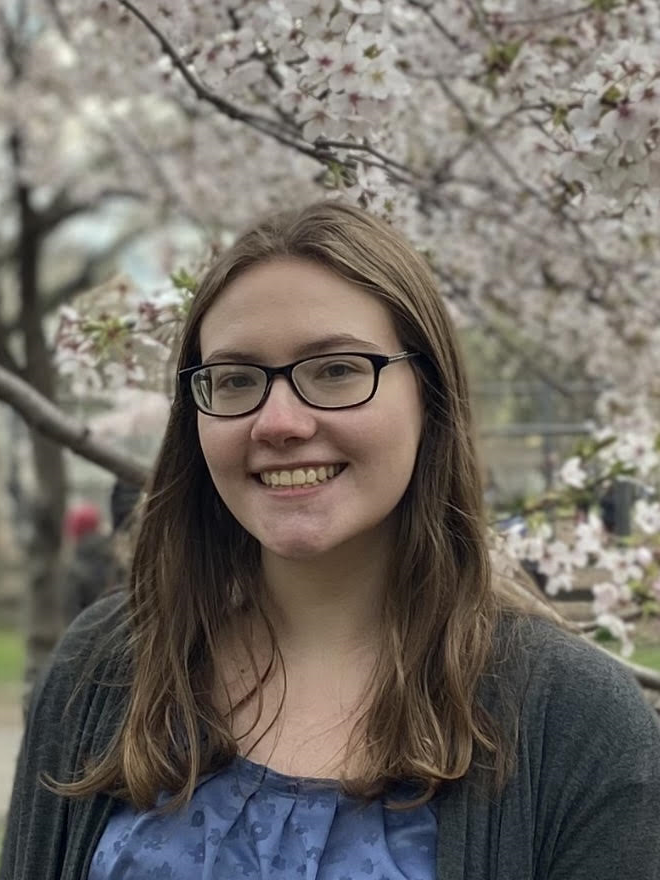}}]
{Erica Heller} received the B.S. degree in physics from Pennsylvania State University, University Park, PA, USA, in 2023.

She is currently pursuing a Ph.D. degree in biomedical engineering at Stony Brook University, Stony Brook, NY, USA. Her research interests include terahertz polarimetric imaging, terahertz imaging of biological materials, and medical imaging. 
\end{IEEEbiography}

\begin{IEEEbiography}
[{\includegraphics[width=1in,height=1.25in,clip,keepaspectratio]{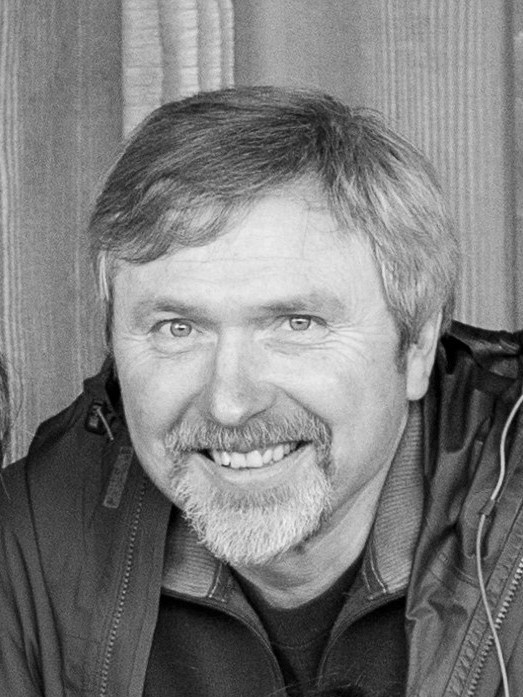}}]
{Paul Vahey} is an Associate Technical Fellow and Analytical Chemist at The Boeing Company. Paul develops nondestructive and  portable spectroscopy methods for process measurements. Prior to Boeing, Paul advanced Process Analytical Technologies in pharmaceutical and chemical industries. Paul earned Chemistry degrees from Harvey Mudd College and the University of Washington.
\end{IEEEbiography}

\begin{IEEEbiography}
[{\includegraphics[width=1in,height=1.25in,clip,keepaspectratio]{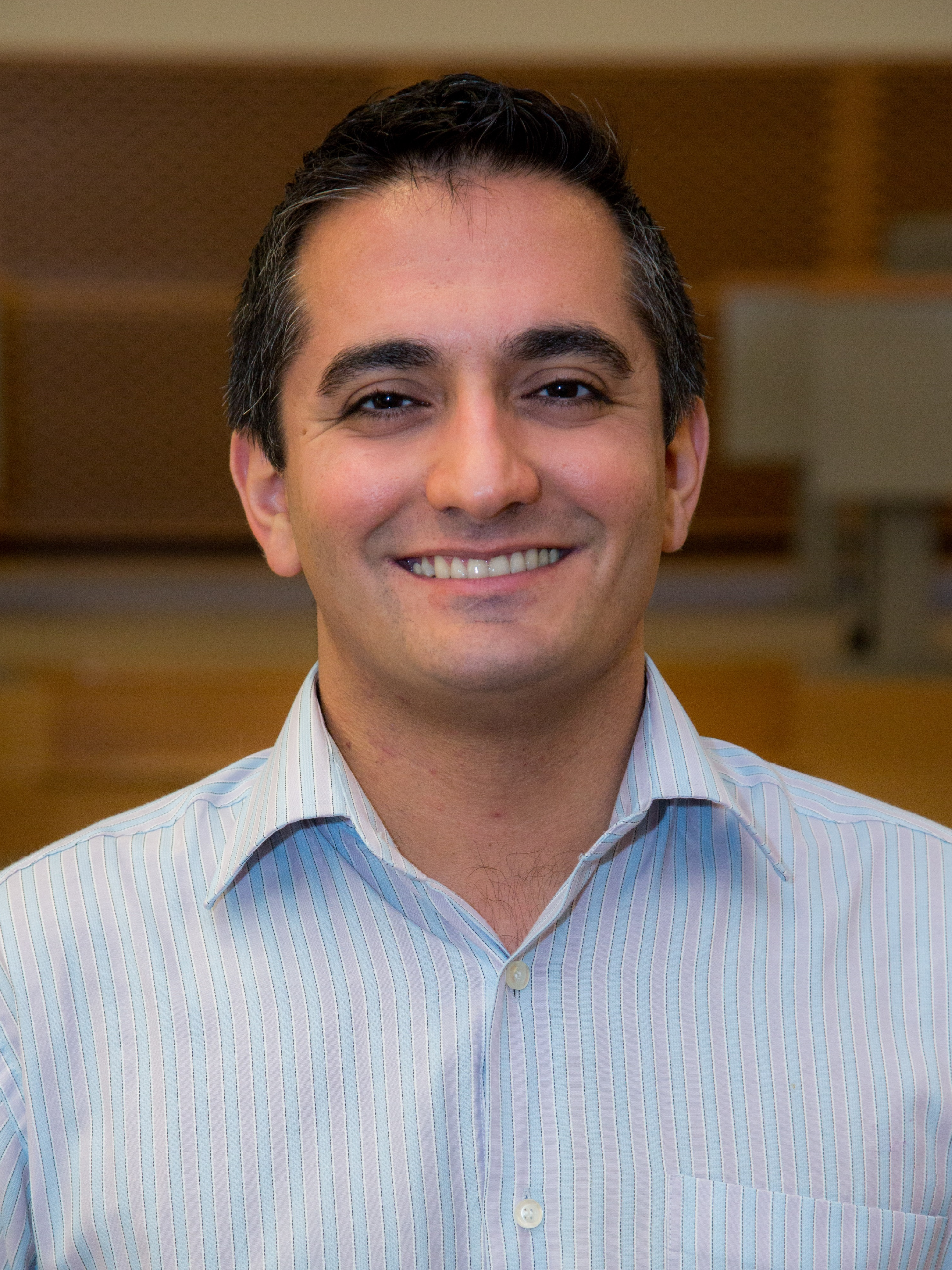}}] {M. Hassan Arbab} is an associate professor of Biomedical Engineering at Stony Brook University. He received the B.S. degree in electrical engineering from Shahid Beheshti University, Tehran, Iran, in 2004, and the M.S. and dual Ph.D. degrees in electrical engineering and nanotechnology from the University of Washington, Seattle, WA, USA, in 2008 and 2012, respectively. He was then awarded the inaugural Applied Physics Laboratory (APL) Director’s Distinguished Post-Doctoral Fellowship. From 2012 to 2016, he was a Postdoctoral Research Associate and a Senior Research Scientist at the Applied Physics Laboratory, University of Washington. He joined Stony Brook University in 2016 as an Assistant Professor and in 2023 became an Associate Professor. His research interests include terahertz science and technology, ultrafast and nonlinear optics, signal and image processing, machine learning, and the biomedical applications of terahertz spectroscopy. He is a member of the IEEE, OPTICA (formerly the Optical Society of America), the International Society for Optical Engineering (SPIE), and the Biomedical Engineering Society.
\end{IEEEbiography}

\vfill

\end{document}